\documentclass{llncs}

\usepackage{gensymb}
\usepackage{enumitem}
\usepackage{tabularx}
\usepackage{url}

\begin{document}


\title{A Multi-modal Deep Neural Network approach to Bird-song identification}

\author{Botond Fazekas\inst{1} \and Alexander Schindler\inst{1,2} \and Thomas Lidy\inst{1} \and Andreas Rauber\inst{1}}
\tocauthor{Botond Fazekas, Alexander Schindler, Thomas Lidy, Andreas Rauber}
\authorrunning{Botond Fazekas et al.} 

\institute{
Vienna University of Technology\\
Institute of Software Technology and Interactive Systems, Vienna, Austria\\
\email{botond.fazekas@tuwien.ac.at, lidy@ifs.tuwien.ac.at},\\ 
\and
Austrian Institute of Technology \\
Center for Digital Safety and Security, Vienna, Austria \\
\email{alexander.schindler@ait.ac.at} }

\maketitle

\begin{abstract}


We present a multi-modal Deep Neural Network (DNN) approach for bird song identification. The presented approach takes both audio samples and metadata as input. The audio is fed into a Convolutional Neural Network (CNN) using four convolutional layers. The additionally provided metadata is processed using fully connected layers. The flattened convolutional layers and the fully connected layer of the metadata are joined and fed into a  fully connected layer.
The resulting architecture achieved 2., 3. and 4. rank in the BirdCLEF2017 task in various training configurations.


\end{abstract}

\section{Approach}

We present our multi-modal Deep Neural Network (DNN) submission to the BirdCLEF 2017 task\cite{birdclef2017} which is part of the LifeCLEF \cite{joly2016} Multimedia Retrieval of biodiversity data evaluation campaign. The presented system is an adoption of the approach introduced in \cite{lidy2016} which was extensively evaluated in \cite{schindler2016}. The presented approach extends the originally audio-only based model to include further modalities as input. The original part is based on the provided samples of field recorded audio content. This information is converted to the frequency domain and sequentially processed before being fed into a custom Convolutional Neural Network (CNN) using four convolutional layers. The additionally provided metadata is processed using fully connected layers. The flattened convolutional layers and the fully connected layer of the metadata were joined and fed into a large fully-connected layer.

To achieve better convergence of the neural networks and to improve model accuracy several pre-processing steps are applied to the provided data. The field-recordings are split into bird-song and noise parts. For the training of the models a random audio segment is selected from the sound file. Various data-augmentation steps detailed below are applied to the Mel-scaled spectrograms and fed into the network. From the provided metadata, longitude, latitude, elevation and the part of the day are used as additional information. Each feature is flagged with an extra bit in case of missing data.

For the final calculation of the results, sequential audio segments with 50\% overlap are taken from the sound files and predictions are retrieved from the trained model. To asses the final classification, the average predictions for each segment of a sound file is calculated.


\section{Preprocessing}

This section describes data transformations, especially in the audio domain, including data manipulation methods to augment the provided training data.

\subsection{Sound preprocessing}

For the sound preprocessing a similar method formulated in \cite{spren:jaggi} is applied. The audio recordings are split in sound, noise and irrelevant segments. In order to do that we compute the spectrogram of the sound file using short-time Fourier transform (STFT) with a Hanning window function (size 512, 74\% overlap). We normalize the resulting spectrogram to the interval [0,1]. The spectrogram is treated then as a grayscale image.

As in most of the recordings the foreground bird singing/calling has higher amplitude than the background noise, in order to distinguish the relevant sound from the background noise, each STFT frequency bin is set to 1 if it is above three times the median of the corresponding row and three times the median of its corresponding column, otherwise it is set to 0. However, as this step results in a noisy spectrogram, binary erosion and dilation filter is applied to it. We have used a 4 by 4 filter as suggested in \cite{spren:jaggi}. A one dimensional indicator vector is created from this image in which the $i$-th column is set to 1 if the corresponding column in the spectrogram contains at least one 1, other it is set to 0. This vector is further binary dilated twice. The indicator vector is then scaled to the original length of the recording and it is used as a mask to extract the relevant sound part.
%
For separating the noise the same method is applied with a threshold of 3 instead of 2.5 for the median clipping and the resulting image is then inverted. Columns containing pixels which don't have an amplitude larger than 3 times or smaller than 2.5 times the row and column median are considered as irrelevant as in this case they cannot be distinguished clearly from the sound or noise part. However, with very noisy recordings or in ones that contain only bird songs without any quiet parts this approach can result in very short or even empty segments, as none or very few pixels will be above the median threshold. To overcome the problem of short segments, a minimum segment length of 32.768 samples is selected, as this is the minimum sound chunk size in our network architecture. The noise/sound separation threshold is iteratively lowered by 0.1 until the length of the sound part is over this limit.

\noindent
Since the Deep Learning network needs a fix sized input during the training, for composing the batches we randomly select 16 (our batch size) files from which we randomly select segments. If the files contain less than 32768 samples, instead of padding we loop the files. The selected segments are then converted to the time-domain using Short-term Fourier Transform (STFT). In a subsequent step a log-normalized Mel-scale Transform with with 80 Mel-bands is applied.

\subsection{Data Augmentation}

Most of the data augmentation steps are similar to the ones used in \cite{spren:jaggi}. However, we found that in addition to the data augmentation steps proposed small variations in the amplitude, overlaying other birds from neighboring areas can further improve the accuracy, leading to the data augmentation process described below.

%
%

\begin{description}[leftmargin=2.6cm,itemsep=3pt,labelwidth=2.4cm,labelindent=0cm,rightmargin=0.1cm] 

\item[Noise overlay] 
During the training up to 4 random noise samples are taken from the noise files of the training set and each is added With 75\% probability to the sound sample. This results in having some segments containing no noise overlay at all, while others having four different. Adding more noise to the sound samples results in a worse performance. We found that greatly dampening the volume of the noise (as described in \cite{spren:jaggi} reduces the accuracy. Thus the overlay volume is only changed by $\pm 10\%$.

\item[Combining same class audio files]
With a probability of 70\%, recordings of birds from the same class are overlayed with a random damping factor between 20\% and 60\%.

\item[Combining birds from the neighboring area]
In addition to the noise, with a probability of 30\%, a bird singing/calling of a different class that can be found in a distance of 1\degree{} East/West/North/South is overlayed on the sample with $30\% \pm 5\%$ damping.

\item[Random cut]
After applying one of the above described overlays the spectrogram is randomly cut into two parts and the two parts concatenated again after switching the order. 
%

\item[Volume shift]
The volume of the input audio is randomly changed by $\pm 5\%$.

\item[Pitch shift]
The pitch of the input audio is randomly changed by $\pm 5\%$.

%
%

\end{description}

\subsection{Metadata preprocessing}

To incorporate the available metadata in the model some preprocessing is required due to missing or inapplicable values. For the missing values we use other instances of the same species where these attributes are available. We calculate the mean and the variance of the respective attribute distribution and generate a normal distributed random value.

\pagebreak

\noindent
Apart from the date and the geo-coded coordinates, the time of the day is available. If this information is missing it is randomly generated as above. It has been shown that bird song intensity correlates with the melatonin levels in the birds and thus with the daylight \cite{ball:balthazart,bentley}. As the time of the sunrise and sunset varies during the year, the time of the day is not directly related to the amount of light. Thus, instead of directly using the time values we decided to divide the day into six different categories of sunlight exposure corresponding to different positions of the sun on the sky. The time of the sunrise and the sunset is dependent on the coordinates and on the day of the year (and partially on the elevation, however this is ignored in our implementation) and it is approximated with the algorithm formulated in \cite{astro}. We define the following parts of a day:

\begin{itemize}
\item Night$_1$ - From midnight until the sun is 9\degree{} below the horizon (BTH)
\item Dawn - From 9\degree{} BTH until 4\degree{} above the horizon (ATH)
\item Forenoon - From 4\degree{} ATH until noon
\item Afternoon - From noon until 4\degree{} ATH
\item Dusk - From 4 \degree{} ATH to 9\degree{} BTH
\item Night$_2$ - 9\degree{} BTH until midnight
\end{itemize}

\noindent
9\degree{} BTH is selected because it lies between the nautical twilight (i.e. the horizon is visible) and the civil twilight (i.e. terrestrial objects are visible to the human eye), 4\degree{} ATH is selected arbitrarily as a point where the sun is already clearly above the horizon.

\section{Network architecture}

The network has two types of input: one for the spectrogram and one for the metadata. The metadata input is a vector of 7 elements:

\begin{enumerate}
\item Coordinates available (1 if available, 0 if not)
\item Latitude (normalized to 0..1)
\item Longitude (normalized to 0..1)
\item Elevation available (1 if available, 0 if not)
\item Elevation (normalized to 0..1)
\item Part of day available  (1 if available, 0 if not)
\item Part of day (normalized to 0..1)
\end{enumerate}

The metadata input is fed into a fully connected layer of 100 neurons. The spectrogram input layer ($80 \times 512$ units) is followed by four convolutional layers with Exponential Linear Unit (ELU) activation, each followed by a max-pooling layer. We found that ELUs yield the same results as using a rectifying activation function but without the need of batch normalization and reducing the training times by about 300\%. A dropout of 0.2 is used on the input layer, after flattening the convolutional layers (0.4) and after the fully connected layer (0.4).

We use either an FFT window of 256 with an 32768 sample long sound segment or an FFT window of 512 with an 65536 sample long sound segment.
%
%
For both of them the Mel scale is calculated with 80 bands. Thus, the input layer is a matrix of $80 \times 512$. It is important to note that having 80 Mel bands with an FFT windows size of 256 means that some Mel bands are empty. However, these are filtered anyway in the first convolutional / max-pooling layer. Thus, instead of using differently configured input layers for FFT sizes 256 and 512 we are using a single input layer configuration covering all bands. 
%
%
%
For the training we use a batch size of 16, and a learning rate of 0.001, with Nesterov momentum of 0.9.

\section{Results}

\begin{description}[leftmargin=3.3cm,itemsep=1pt,labelwidth=3.1cm,labelindent=0cm,rightmargin=0.1cm] 

\item [Cynapse Run 1:]
In this run we use an 256 FFT window, and we train the network with 90\% of the training set for 4 days. 

\item [Cynapse Run 2:]
We use an 512 FFT window, and the training is kept running for 3 days with 90\% of the training set. Then, for 1 day it is trained on the whole training set.

\item [Cynapse Run 3:]
The network from Cynapse Run 1 is kept training for an additional day with the whole training set.

\item [Cynapse Run 4:]
The predictions of Cynapse Run 2 and Cynapse Run 3 were taken and averaged for each class.
\end{description}


\newcolumntype{Y}{>{\centering\arraybackslash}X}

\begin{table}[]
\centering
\caption{Results of the Cynapse Runs (CR) on the four BirdCLEF evaluation tasks: MAP '16 Soundscapes with time-codes (MAP w TC), MAP Soundscapes without time-codes (MAP wo TC), MAP Traditional records only main species (MAP o MS), MAP Traditional records with background species (MAP w BS)}
\label{my-label}
\begin{tabularx}{\textwidth}{Y|Y|Y|Y|Y}

\hline
\textbf{CR} & \textbf{MAP w TC} & \textbf{MAP wo TC} & \textbf{MAP o MS} & \textbf{MAP w BS} \\
\hline
1  & 0,165          & 0,008           & 0,486          & 0,432          \\
2  & 0,069          & \textbf{0,012}  & 0,562          & 0,493          \\
3  & \textbf{0,168} & 0,008           & 0,514          & 0,456          \\
4  & 0,142          & 0,010           & \textbf{0,579} & \textbf{0,511} \\
\hline
\end{tabularx}
\end{table}

For detailed scenario descriptions and a full report on the BirdCLEF evaluation campaign results, please refer to the BirdCLEF 2017 Web-page\footnote{\url{http://www.imageclef.org/lifeclef/2017/bird}}. We also tested a more complicated architecture with three inputs: an FFT 256 spectrogram, an FFT 512 spectrogram and the metadata which are then co-joined in a fully connected layer. This architecture yielded better results as the FFT 512-only network, but worse than the FFT-256 network, however the training time is considerably longer.

\section{Conclusions and Future Work}

The presented approach harnesses information deriving from multiple modalities. 
%
Cynapse Run 3 performed the best for the time-coded soundscapes, which may contain longer parts without any relevant sound. However for the traditional records with less noise it had the worst performance.
On the other hand, Cynapse Run 4 that incorporated the results of Cynapse Run 3 and the Cynapse Run 2 results with higher frequency resolutions performed the best for traditional record, but only third best for the time-coded soundscapes.
A possible explanation could be that the higher frequency resolution is a more distinguishing feature than the higher temporal resolution but only if there are long enough sound segments available.
%
Future work would focus on regarding ornithological relationships instead of treating each species as an isolated class.




\begin{thebibliography}{5}
%
\bibitem{birdclef2017}
Go{\"e}au, Herv{\'e} and Glotin, Herv{\'e} and Planqu\'e, Robert and Vellinga, Willem-Pier and Joly, Alexis
LifeCLEF Bird Identification Task 2017
CLEF working notes 2017


\bibitem {spren:jaggi}
Elias Sprengel, Martin Jaggi, Yannic Kilcher, and Thomas Hofmann (2016)
Audio Based Bird Species Identification using Deep Learning Techniques.
CLEF 2016 Working Notes

\bibitem {ball:balthazart}
Ball, G.F. \& Balthazart, J. (2002)
Neuroendocrine mechanisms regulating reproductive cycles and reproductive behavior in birds.
In Hormones, Brain, and Behavior. 2:649–798

\bibitem {bentley}
Bentley, G.E.; Van’t Hof, T.J.; Ball, G.F. (1999).
Seasonal neuroplasticity in the songbird telencephalon: A role for melatonin.
In Proceedings of the National Academy of Sciences of the United States of America.
Proceedings of the National Academy of Sciences 96(8):4674-4679

\bibitem{astro}
Van Flandern, T. C., and K. F. Pulkkinen. (1979).
Low-precision formulae for planetary positions. 
In The Astrophysical Journal Supplement Series 41:391-411.
APA

\bibitem{joly2016}
Joly, A., Goëau, H., Glotin, H., Spampinato, C., Bonnet, P., Vellinga, W. P., Müller, H. (2016). 
LifeCLEF 2016: mMltimedia life species identification challenges. 
In International Conference of the Cross-Language Evaluation Forum for European Languages (pp. 286-310). Springer.

\bibitem{schindler2016}
A. Schindler, T. Lidy, A. Rauber. (2016)
Comparing shallow versus deep neural network architectures for automatic music genre classification. 
In Proceedings of the 9th Forum Media Technology (FMT2016), St. Poelten, Austria.

\bibitem{lidy2016}
T. Lidy, A. Schindler. (2016).
CQT-based convolutional neural networks for audio scene classification. 
In Proceedings of the Detection and Classification of Acoustic Scenes and Events 2016 Workshop (DCASE2016), (pages 60--64).

\end{thebibliography}
\end{document}